\documentclass[conference]{IEEEtran}
\IEEEoverridecommandlockouts

\usepackage{cite}
\usepackage{amsmath,amssymb,amsfonts}
\usepackage{algorithmic}
\usepackage{graphicx}
\usepackage{textcomp}
\usepackage{xcolor}
\usepackage{booktabs}
\usepackage{listings}
\usepackage[hidelinks]{hyperref}

\lstdefinestyle{ieeecode}{
  basicstyle=\ttfamily\footnotesize,
  numbers=left,
  numberstyle=\tiny,
  stepnumber=1,
  numbersep=5pt,
  breaklines=true,
  breakatwhitespace=false,
  frame=single,
  tabsize=2,
  captionpos=b,
  keywordstyle=\color{blue},
  commentstyle=\color{gray},
  stringstyle=\color{red},
  showstringspaces=false,
  language=Python
}

\def\BibTeX{{\rm B\kern-.05em{\sc i\kern-.025em b}\kern-.08em
    T\kern-.1667em\lower.7ex\hbox{E}\kern-.125emX}}
\begin{document}

\title{Inferring Attributed Grammars from Parser Implementations}


\author{\IEEEauthorblockN{Andreas Pointner}
\IEEEauthorblockA{Advanced Information Systems and Technology\\University of Applied Sciences\\Upper Austria, Campus Hagenberg\\
Softwarepark 11\\
4232 Hagenberg i. M., Austria\\
Email: andreas.pointner@fh-hagenberg.at}
\and
\IEEEauthorblockN{Josef Pichler}
\IEEEauthorblockA{Department for Software Engineering\\University of Applied Sciences\\Upper Austria, Campus Hagenberg\\
Softwarepark 11\\
4232 Hagenberg i. M., Austria\\
Email: josef.pichler@fh-hagenberg.at}
\and
\IEEEauthorblockN{Herbert Pr{\"a}hofer}
\IEEEauthorblockA{Institut for System Software\\Johannes Kepler University Linz\\
Altenberger Stra{\ss}e 69\\
4040 Linz, Austria\\
Email: herbert.praehofer@jku.at}}

\maketitle

\begin{abstract}
Software systems that process structured inputs often lack complete and up-to-date specifications, which specify the input syntax and the semantics of input processing. While grammar mining techniques have focused on recovering syntactic structures, the semantics of input processing remains largely unexplored. In this work, we introduce a novel approach for inferring attributed grammars from parser implementations. Given an input grammar, our technique dynamically analyzes the implementation of recursive descent parsers to reconstruct the semantic aspects of input handling, resulting in specifications in the form of attributed grammars. By observing program executions and mapping the program’s runtime behavior to the grammar, we systematically extract and embed semantic actions into the grammar rules. This enables comprehensive specification recovery. We demonstrate the feasibility of our approach using an initial set of programs, showing that it can accurately reproduce program behavior through the generated attributed grammars.
\end{abstract}


\section{Introduction}

Software systems that process structured input such as configuration files, domain-specific languages, or communication protocols typically use parsers based on context-free grammars to define valid syntax. However, real-world parsers also perform semantic tasks such as value extraction, state updates, or transformations, which are often embedded in imperative code~\cite{Deursen1998, schroder_grammars_2022}.

Attributed grammars~\cite{knuth_semantics_1968} combine syntax and semantics in a single formalism. They extend context-free grammars by attaching attributes and semantic rules to grammar productions. As such, they serve as executable specifications that describe both how input is recognized and how it is processed. Attributed grammars can therefore act as precise and formal specifications for parser programs.

For many existing systems, such attributed grammars are not available. While some efforts exist to recover syntactic grammars through grammar mining techniques~\cite{stevenson_survey_2014}, these focus only on reconstructing the structural part of the input language. Grammar mining is a well-studied problem and assumes that parsing follows a certain style, most commonly recursive descent parsing. Tools like Mimid~\cite{gopinath_mining_2020} and Autogram~\cite{hoschele_mining_2016} have demonstrated that such techniques can extract context-free grammars from real software.

Our approach builds on existing syntactic grammars, assuming a grammar is already available. We introduce a novel method to automatically infer attributed grammars from recursive descent parsers by analyzing their implementation, executing them on generated inputs, and mapping runtime behavior to grammar elements. This yields complete attributed grammars that capture both syntax and semantics, enabling formally defined executable specifications where they were previously absent.

The concepts of grammar mining, attribute grammar mining, and grammar fuzzing were initially introduced in our earlier works \cite{moser_towards_2022, pointner_mining_2023}. In a subsequent work \cite{pointner_generating_2025}, we demonstrated enhancements to the grammar mining process. Building upon these foundations, the present work reports on preliminary results in the direction of attribute grammar mining.


In the next section, we review related work before describing our approach in \autoref{sec:approach}. \autoref{sec:results} contains a preliminary evaluation and initial findings. \autoref{sec:summary} concludes with a summary and outlook on future work.  

\section{Related Work}

A significant body of research aims to extract the semantic information of program executions. For instance, Kalita~et~al.~\cite{kalita_synthesis_2022} present an approach to synthesizing semantic actions in attributed grammars using input-output examples and constraint-based program synthesis. Their method assumes a partially annotated grammar and focuses on filling in missing semantic actions through symbolic evaluation.

In a similar spirit of inferring semantics from observable behavior, the work of Chen~et~al.~\cite{chen_towards_2017} describe a neural program synthesis approach that learns to generate parsers from input-output pairs using reinforcement learning. Their method focuses on inferring parser behavior from data without access to code or grammar.

Complementing these data-driven approaches, Steinhöfel and~Zeller~\cite{steinhofel_input_2022} present ISLa, a declarative specification language that augments context-free grammars with semantic constraints to generate and validate semantically valid structured inputs. They also introduce ISLearn, a tool that mines such constraints from sample inputs using pattern-based abstraction and validation.

Taking a different direction, Kova\u{c}evi'{c}~et~al.~\cite{kovacevic_grammar_2020} propose an evolutionary approach to semantic inference using genetic programming to infer attributed grammars from sample programs annotated with semantic meanings. Their tool LISA.SI evolves candidate semantic rules based on fitness evaluations against example outputs, starting from a predefined grammar.


In addition to the presented state of the art, there is ongoing work by Mera~\cite{mera_mining_2019} who proposes the idea of extracting constraints and semantics from programs to enrich grammars. However, this work remains at an early conceptual stage and does not provide concrete results or detailed descriptions of the approach.

In contrast to prior work, our approach systematically recovers full attributed grammars through dynamic analysis and trace-based mapping. Rather than relying on partially annotated grammars, training data, or manual labeling, it extracts semantic information directly from program implementations. This enables robust mappings between code and grammar structures and transforms syntactic grammars into comprehensive attributed grammars, supporting full specification recovery beyond input generation or parser learning.

\section{Approach}\label{sec:approach}

Our approach targets software systems or components that process structured inputs. We consider a program \textbf{P} that takes an input \textbf{I} and produces an output \textbf{O}: $I \rightarrow P \rightarrow O$.

The output may be the final result or an intermediate representation, such as an in-memory data structure (e.g., a JSON dictionary). 

Such functionality is commonly implemented by dedicated parsers. The goal of our work is to reconstruct an attributed grammar \textbf{AG} for the program \textbf{P} that captures both the syntactic structure of valid inputs and their semantic transformation into outputs. In this way, AG becomes a full specification of the program’s input-processing behavior: $I \rightarrow AG \rightarrow O$.

We build on the fact that in recursive-descent parsers, syntax and semantics are often implemented together. Each function typically corresponds to a grammar rule and processes part of the input while constructing the semantic output. Thus, for reconstructing an attributed grammar from a program, we must distinguish between statements responsible for syntax analysis and those performing semantic processing. 

To illustrate our approach, we use a simplified number parser shown in \autoref{lst:sample} as a running example. The parser accepts positive and negative integers from an input string and returns the corresponding Python \textit{integer} value. It exemplifies the typical integration of syntax analysis and semantic processing in recursive descent parsers, checking for valid structure while constructing the output value through embedded semantic actions.

\begin{lstlisting}[style=ieeecode, float=ht, caption={Parser methods with intertwined syntax analysis (Syn) and semantic actions
(Sem).}, label={lst:sample}]
s = '...'                              # Given Input
i = 0

def number():
  sign = 1                                     # Sem
  if s[i] == '-':                              # Syn
    sign = -1                                  # Sem
    i += 1                                     # Syn
  if s[i] in '0123456789':                     # Syn
    value = digit()                            # Sem
    while s[i] in '0123456789':                # Syn
      value = 10 * value + digit()             # Sem
    return sign * value                        # Sem
  else:                                        # Syn
    raise ParserException("Expected a number") # Syn

def digit():
  if s[i] in '0123456789':                     # Syn
    res = int(s[i])                            # Sem
    i += 1                                     # Syn
    return res                                 # Sem
  else:                                        # Syn
    raise ParserException("Expected a digit")  # Syn
\end{lstlisting}

Our approach assumes the existence of a syntactic grammar, which may have been previously mined using techniques such as Mimid~\cite{gopinath_mining_2020}. 
For the program in \autoref{lst:sample}, we assume that a grammar as shown below has already been mined. The grammar allows an optional minus sign followed by one or more digits. 

\begin{lstlisting}[basicstyle=\ttfamily\footnotesize]
NumberParser = ['-'] Digit { Digit } .
Digit = '0' | '1' |... | '9' .
\end{lstlisting}

Our goal is to enrich the syntactic grammar with semantic rules, resulting in an attributed grammar (AG). The AG uses synthesized attributes (marked with $\uparrow$) to propagate semantic information, defined within \texttt{sem ... endsem} blocks. For example, each \texttt{Digit} rule assigns a numeric value to attribute \texttt{D}, while the \texttt{NumberParser} rule computes the final value \texttt{N} by aggregating digits and handling a leading minus sign. These semantics incrementally construct the integer value during parsing.

\begin{lstlisting}[basicstyle=\ttfamily\footnotesize, mathescape=true]
Grammar(NumberParser):
NumberParser$\uparrow$N = 
             sem sign = 1 endsem
  [ '-'      sem sign = -1 endsem
  ]
  Digit$\uparrow$D     sem value = D endsem
  { Digit$\uparrow$D2  sem value = (10 * value) + D2 endsem
  }          sem N = (sign * value) endsem .
Digit$\uparrow$D = 
 '0'         sem D = int('0') endsem
 | '1'       sem D = int('1') endsem
 ...
 | '9'       sem D = int('9') endsem .
\end{lstlisting}

\begin{figure}[htb]
    \centering
    \includegraphics[width=\linewidth]{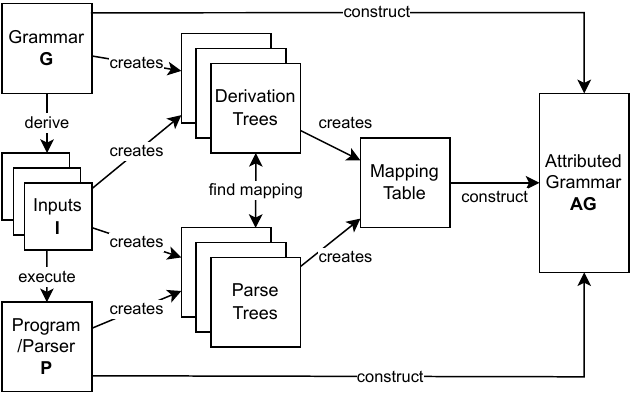}
    \caption{Overview of the approach: parse trees are created from program execution and derivation trees from grammar parsing, using the same input. A mapping between both trees enables the transfer of semantic information.}
    \label{fig:overview}
\end{figure}

\autoref{fig:overview} illustrates our approach. Given a grammar $G$ and a program $P$, we generate a set of syntactically valid inputs $I$ from $G$. Each input $i \in I$ is executed by $P$ to produce parse trees reflecting its runtime behavior, while derivation trees are constructed from $G$. These trees are structurally matched to create a mapping between grammar constructs and program operations. Based on this mapping, we enrich $G$ with inferred semantic rules, resulting in an attributed grammar $AG$ that captures both syntax and semantics.

Thus, the approach consists of the following steps: (1) input generation, (2) construction of the parse tree, (3) construction of the derivation tree, (4) mapping of tree nodes, and (5) transfer of semantic actions. The steps are described in detail in the following subsections. 

\subsection{Input Generation}

Program inputs are crucial for generating parse and derivation trees. To enable complete semantic mapping, the input set must provide sufficient coverage of the grammar, including all terminal and non-terminal symbols, production rules, and positions within rules. This ensures that each grammar element's position is exercised by at least one input, allowing accurate alignment with program behavior and correct extraction of semantic actions.

To achieve the required grammar coverage, we apply grammar fuzzing techniques that combine the Efficient Grammar Fuzzer~\cite{zeller_fuzzing_2019} with coverage based strategies from Grammarinator~\cite{hodovan_grammarinator_2018}. This combination enables systematic exploration of the grammar. The Efficient Grammar Fuzzer steers input generation toward untested features using feedback, while Grammarinator’s cooldown strategy reduces repetition by penalizing frequently used rules. Together, they ensure broad and balanced grammar coverage without manual input crafting.

In our running example, the inputs \texttt{-74521}, \texttt{-9836}, and \texttt{0686} cover the entire grammar. They include the optional minus sign, repeated digits, and all digits from 0 to 9, exercising optional elements, repetitions, and structural variations. Although the grammar defines infinitely many valid inputs, this set provides full rule coverage for semantic extraction.

\subsection{Derivation Tree Construction}

\begin{figure}[htb]
    \centering
    \includegraphics[width=\linewidth]{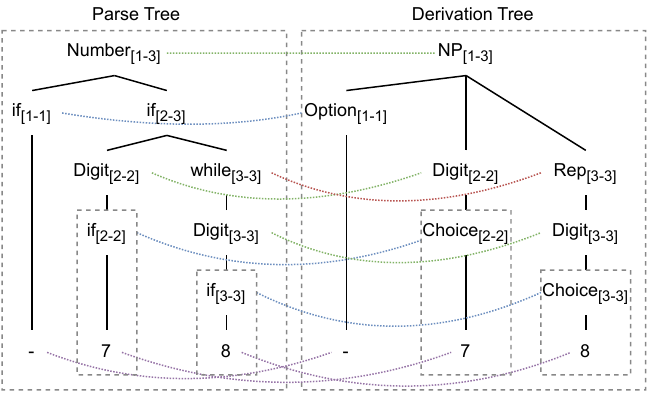}
    \caption{The parse tree on the left side and the derivation tree on the right side. The subscript numbers of the nodes, represent the terminal yield. Dotted connection lines, show the nodes that can be mapped with each other. Inner boxes show the grammar rule and function scope boundaries.}
    \label{fig:trees}
\end{figure}

Derivation tree construction from a grammar G and a given input I works by parsing the input I from left to right and applying left canonical derivations starting from the start symbol~\cite{aho_compilers_2014}. The right tree shown in \autoref{fig:trees} shows the derivation tree for input \texttt{-78}. For non-terminal symbols or options, choices, or repetitions, an inner node is created and each terminal symbol results in a leaf node. Thus, each node in the derivation tree has a type as shown in the right column in \autoref{tab:node-types}: \textit{Terminal Symbol}, \textit{Non-Terminal Symbol}, \textit{Repetition}, \textit{Option} or \textit{Choice}.
 
Additionally, a \textit{terminal yield} is maintained with inner nodes in the tree, which refers to the sequence of input symbols that the corresponding grammar element consumes during parsing. We represent this yield as a position range over the input, which is shown as a subscript on the node in the tree. For example, the terminal yield of the \textit{Option} node is the minus sign, corresponding to the input range [1–1], while the root node covers the entire input -78 with the range [1–3]. 


\subsection{Parse Tree Construction}

Parse tree construction from program P and a given set of inputs follows the approach introduced by Mimid \cite{gopinath_mining_2020}. When executing P with a specific input I, the following observations are made during runtime:

\begin{enumerate}
    \item Accesses to input characters are interpreted as terminal symbols in the grammar.
    \item Method or function calls represent non-terminal symbols and indicate the application of grammar rules.
    \item Control flow statements such as if or case reflect alternatives or optional grammar elements.
    \item Loop constructs are associated with repetition and are used to build repeating elements in the grammar.
\end{enumerate}

These observations are used to dynamically construct a parse tree that reflects how the program processes the input. The left tree shown in \autoref{fig:trees} shows the parse tree generated by executing program P (shown in \autoref{lst:sample}) with the input \texttt{-78}. As the program runs, each relevant execution step contributes a node to the parse tree. For example, when the number function (line 1) is called, the root node Number\textsubscript{[1-3]} is created. The conditional in line 3, which checks for a minus sign, results in the node if\textsubscript{[1-1]}. Subsequent calls to the digit function in line 7 generate the node Digit\textsubscript{[2-2]}. Similarly, the while loop in line 8 leads to the creation of the node while\textsubscript{[3-3]}.

As in the derivation tree, nodes in the parse tree are assigned specific types: \textit{Input Characters}, \textit{Functions}, \textit{Loops}, and \textit{Conditions}. These are shown in \autoref{tab:node-types}. Moreover, as in the derivation tree each inner node has the terminal yield range assigned to it, shown as subscripts in \autoref{tab:node-types}. In the parse tree, the terminal yield represents the part of the input parsed by that part of the program. 


\subsection{Tree Node Mapping}

The core idea behind the tree node mapping is to match nodes based on their types, as defined in \autoref{tab:node-types}, and their terminal yield ranges, indicated by subscript numbers in \autoref{fig:trees}. As indicated in \autoref{tab:node-types}, we assume that Input Character nodes in the parse tree correspond to Terminal nodes in the derivation tree. Likewise, function nodes correspond to non-terminal symbols while loops and conditions correspond to repetitions and options or choices, respectively. In \autoref{fig:trees}, the result of this mapping is visualized with dotted lines.

\begin{table}[htb]
\caption{Node types used for mapping, with examples from the parse tree and derivation tree shown in \autoref{fig:trees}}
\label{tab:node-types}
\centering
\begin{tabular}{ll}
\toprule
Parse Tree & Derivation Tree \\ \midrule
Input Characters (-, 7, 8) & Terminal Symbols (-, 7, 8) \\
\begin{tabular}[c]{@{}l@{}}Functions (Number\textsubscript{[1-3]}, \\ ~~~~~~~~Digit\textsubscript{[2-2]}, Digit\textsubscript{[3-3]})\end{tabular} & \begin{tabular}[c]{@{}l@{}}Non-Terminal Symbol (NP\textsubscript{[1-3]}, \\ ~~~~~~~~~Digit\textsubscript{[2-2]}, Digit\textsubscript{[3-3]})\end{tabular} \\
Loops (while\textsubscript{[3-3]}) & Repetitions (Rep\textsubscript{[3-3]}) \\
Conditions (if\textsubscript{[1-1]}) & Options (Option\textsubscript{[1-1]}) \\
Conditions (if\textsubscript{[2-2]}, if\textsubscript{[3-3]}) & Choices (Choice\textsubscript{[2-2]}, Choice\textsubscript{[3-3]}) \\ \bottomrule
\end{tabular}
\end{table}

While the mapping is straightforward when both trees are structurally similar, we also account for minor differences. These include intermediate conditional nodes or slight variations in terminal yield ranges, such as the node if\textsubscript{[2-3]} in the parse tree. 

To handle such cases, we perform our mapping in four steps. First we map input characters to terminals. Following that, functions and non-terminal symbols are mapped. Next, control-flow nodes and grammar structures are mapped within the boundaries of the already mapped nodes. Finally, if unmapped nodes remain their semantics are moved to suitable positions. This process is described in detail in the next sections.

\subsubsection{Mapping Input Characters and Terminals}

Input characters and terminal symbols are mapped using in-order depth-first traversal of the parse and derivation trees, producing matching terminal sequences. Since both trees originate from the same input, these sequences must align exactly. Any discrepancy causes the mapping to fail and the process to be aborted. This step ensures alignment of the terminal yields from both trees.

\subsubsection{Mapping Functions and Non-Terminal Symbols}

The mapping between non-terminal symbols in the grammar and parser function calls in the parse tree is performed in a bottom-up manner. Starting from terminal symbols that are already mapped to the derivation tree, the algorithm traverses upward toward the root. It looks for the first pair of nodes in both trees with identical terminal yields and maps them. If the yields differ, the algorithm continues upward in the tree that covers the smaller range. This ensures that each parser function is matched with the most specific corresponding grammar rule.

In our running example, the algorithm begins at the terminal symbol for the minus sign and ascends both trees until it maps Number\textsubscript{[1-3]} in the parse tree to NP\textsubscript{[1-3]} in the derivation tree. Since these are the root nodes, it continues with the next terminal symbol, digit one, and maps Digit\textsubscript{[2-2]} in both trees. It then revisits Number and NumberParse, but since they were already processed, the algorithm proceeds to the final terminal, digit two, completing the mapping at Digit\textsubscript{[3-3]}.

\subsubsection{Mapping of Control-Flow Nodes and Grammar Structures}

Control flow nodes such as loops and conditionals are mapped to grammar constructs like repetitions, options, and choices using the same type and terminal yield criteria. This mapping is restricted to the local scope of the corresponding function or grammar rule to maintain structural consistency. This is annotated in \autoref{fig:trees} with the dashed boxes.

In our running example, the algorithm maps the control flow node if\textsubscript{[1-1]} to Option\textsubscript{[1-1]}, followed by if\textsubscript{[2-2]} to Choice\textsubscript{[2-2]}, if\textsubscript{[3-3]} to Choice\textsubscript{[3-3]}, and while\textsubscript{[3-3]} to Rep\textsubscript{[3-3]}.

Notably, the node if\textsubscript{[2-3]} does not match any grammar construct. Such unmatched nodes are handled separately by transferring their semantics in the next step.

\subsubsection{Handling of unmapped nodes}

If a node cannot be mapped, execution continues by reassigning its semantics to nearby nodes, specifically its children. Semantics that appear before the node are moved to the first child, and those that appear after it are moved to the last. The relative execution order is preserved, ensuring that semantics remain aligned with the corresponding input segments. For the node if\textsubscript{[2-3]}, this means that its pre-semantics are moved to Digit\textsubscript{[2-2]}, while its post-semantics are moved to Digit\textsubscript{[3-3]}.

\subsection{Transfer of Semantic Actions}

All code elements that appear between the blocks responsible for syntax analysis are considered semantic. These semantic elements are grouped into blocks and placed at the appropriate positions in the grammar according to the established mapping between parse tree and derivation tree nodes. \autoref{tab:semantics-transfer} summarizes how different types of code elements are translated into semantic actions within the grammar.

For our running example from \autoref{lst:sample}, this means that expressions such as \texttt{sign=-1} in line 4 are directly transferred to the corresponding grammar position. Whilst statements like \texttt{value=digit()} in line 7 are translated into the out-attribute \texttt{D} and the semantic rule \texttt{value=D}.

\begin{table}
\centering
\caption{Transfer of Program Statement to Grammar Elements.}
\label{tab:semantics-transfer}
\begin{tabular}{@{}ll@{}}
\toprule
Program Statement & Grammar Element \\ \midrule
Assignment Statement & Assignment Semantic \\
Expression Statement & Expression Semantic \\
Return Statement & Assignment of Out-Attribute \\
Function Call & Usage of Out-Attributes \\
Function Parameters & Assignment and usage of In-Attributes \\ \bottomrule
\end{tabular}
\end{table}

After all steps are completed, the result is an attributed grammar. It combines the original syntax rules with semantic actions taken from the program, placed according to the established mapping. This grammar now represents the specification of the given program P.

\section{Preliminary Evaluation and Results}\label{sec:results}

We have done a preliminary evaluation of our current implementation of the approach based on a variety of examples from the literature. These examples are shown and explained in \autoref{tab:evaluation}.

To evaluate correctness, we run both the original program and an interpreter for the reconstructed attributed grammar on the same inputs and compare their serialized outputs. Accuracy is measured as the proportion of matching outputs, indicating how well the grammar replicates program behavior. For the initial experiments, we manually crafted inputs to achieve full statement coverage and maintain control over evaluation. In future work, we plan to use fuzzing to automatically generate high-coverage inputs for larger examples.

\begin{table}
\centering
\caption{The evaluation examples with a short description and which semantics are valid and invalid}
\label{tab:evaluation}
\resizebox{\columnwidth}{!}{%
\begin{tabular}{@{}lll@{}}
\toprule
Program & Description & Semantic Result \\ \midrule
Calc~\cite{gopinath_mining_2020} & \begin{tabular}[c]{@{}l@{}}Simple Parser for \\ mathematical\\ expressions\end{tabular} & \begin{tabular}[c]{@{}l@{}}Creates data structure\\ representing operators\\ with nested lists\end{tabular} \\
CgiDecode~\cite{gopinath_mining_2020} & \begin{tabular}[c]{@{}l@{}}A simple decode of \\ CGI-encoded strings\end{tabular} & Decoded string value \\
MathExpr~\cite{gopinath_mining_2020} & \begin{tabular}[c]{@{}l@{}}A parser for more complex \\ mathematical expression  \\ incl. mathematical functions\end{tabular} & \begin{tabular}[c]{@{}l@{}}Calculate the mathematical\\ result for the expressions\\ and functions\end{tabular} \\
MicroJson~\cite{gopinath_mining_2020} & Simple parser for JSON & \begin{tabular}[c]{@{}l@{}}Builds the data structure\\ that represents the JSON\\ (e.g. List / Object / ...)\end{tabular} \\
XMLParser~\cite{steinhofel_input_2022} & \begin{tabular}[c]{@{}l@{}}Parser for simple XML \\ structures\end{tabular} & \begin{tabular}[c]{@{}l@{}}Boolean value indicating\\ if all opening and closing\\ tags have same IDs\end{tabular} \\
SimpleExpr~\cite{moser_towards_2022} & \begin{tabular}[c]{@{}l@{}}Parser for simple math \\ expression\end{tabular} & \begin{tabular}[c]{@{}l@{}}Calculate the mathematical\\ result for the expressions\end{tabular} \\
ABC Parser~\cite{kovacevic_grammar_2020} & \begin{tabular}[c]{@{}l@{}}Parser of sequences of \\ $A^{n}B^{n}C^{n}$\end{tabular} & \begin{tabular}[c]{@{}l@{}}Boolean value indicating\\ if As, Bs and Cs occur\\ equally often\end{tabular} \\ \bottomrule
\end{tabular}
}
\end{table}

\begin{table}
\centering
\caption{Results of the semantic grammar enrichment process}
\label{tab:results}
\begin{tabular}{@{}lrrr@{}}
\toprule
Program & \multicolumn{1}{l}{Nr. of Inputs} & \multicolumn{1}{l}{Correct} & \multicolumn{1}{l}{Accuracy} \\ \midrule
Calc~\cite{gopinath_mining_2020} & 10 & 10 & 100\% \\
CgiDecode~\cite{gopinath_mining_2020} & 6 & 6 & 100\% \\
Mathexpr~\cite{gopinath_mining_2020} & 32 & 19 & 59\% \\
MicroJson~\cite{gopinath_mining_2020} & 22 & 13 & 59\% \\
XMLParser~\cite{steinhofel_input_2022} & 5 & 5 & 100\% \\
SimpleExpr~\cite{moser_towards_2022} & 6 & 5 & 83\% \\
ABC Parser~\cite{kovacevic_grammar_2020} & 5 & 5 & 100\%\\ \bottomrule
\end{tabular}
\end{table}

The results in \autoref{tab:results} show that the extracted semantics generally align with the expected specifications. In most cases, the approach successfully derived attributed grammars that enrich the syntax with relevant semantic actions. While three examples did not achieve full coverage, the remaining evaluations demonstrate the method's robustness and applicability. Specifically, for SimpleExpr, the approach fails to capture the division-by-zero exception. In MathExpr, certain functions were not fully covered, and the division-by-zero case remained unhandled. In MicroJson, an incorrect placement of a semantic rule led to inaccuracies. These limitations highlight areas for further refinement.

A direct comparison with existing approaches is only partially possible. The first four benchmarks (Calc, CgiDecode, Mathexpr, MicroJson) are not evaluated in terms of attribute grammars in prior work. SimpleExpr originates from our previous work and consequently yields identical results. For XMLParser, the extracted semantics follow a different methodology but result in similar meaning. Finally, the result for ABC Parser aligns closely with the outcomes reported in existing grammar mining approach, indicating the competitiveness of our method.

\section{Conclusion and Outlook}\label{sec:summary}

Our preliminary evaluation shows that the approach can successfully derive attributed grammars from parser programs. However, it currently captures only semantics along the happy path, meaning executions that complete without exceptions. As a result, semantics related to exceptional or error-handling behavior are not tracked. Since we rely on automatically generated inputs, the effectiveness of the analysis also depends on the quality of these inputs. While they typically suffice to cover syntactic paths, certain behaviors require inputs that meet specific semantic constraints. For example, evaluating $asin(x)$ requires $x \in [-1, 1]$ to avoid runtime exceptions.

In terms of mapping, the approach supports only minor differences between the parse tree and the derivation tree. It does not handle substantial deviations such as grammar rule indirections, function calls, or major control flow differences. For example, the mapping may fail if a grammar uses recursion but the program implements the same logic iteratively. The method is also limited to programs that strictly follow a recursive descent parser structure. Consequently, some examples from prior work had to be adapted to fit this format.

The focus of our future work will be to address such cases, enabling the mapping process to handle larger discrepancies between the parser and the program and ensuring that paths ending with an exception can be extracted correctly.

\section*{Acknowledgment} The research reported in this work has been partly funded by BMIMI, BMWET, and the State of Upper Austria in the frame of the SCCH competence center INTEGRATE (FFG grant no. 892418) part of the FFG COMET Competence Centers for Excellent Technologies Programme. We also gratefully acknowledge funding from the Austrian Research Promotion Agency (FFG) and the University of Applied Sciences Upper Austria as part of the project AG-Fuzzer (FFG grant no. 895972).

\bibliographystyle{IEEEtran}
\bibliography{IEEEabrv,mybibfile}

\begin{thebibliography}{10}
\providecommand{\url}[1]{#1}
\csname url@samestyle\endcsname
\providecommand{\newblock}{\relax}
\providecommand{\bibinfo}[2]{#2}
\providecommand{\BIBentrySTDinterwordspacing}{\spaceskip=0pt\relax}
\providecommand{\BIBentryALTinterwordstretchfactor}{4}
\providecommand{\BIBentryALTinterwordspacing}{\spaceskip=\fontdimen2\font plus
\BIBentryALTinterwordstretchfactor\fontdimen3\font minus \fontdimen4\font\relax}
\providecommand{\BIBforeignlanguage}[2]{{%
\expandafter\ifx\csname l@#1\endcsname\relax
\typeout{** WARNING: IEEEtran.bst: No hyphenation pattern has been}%
\typeout{** loaded for the language `#1'. Using the pattern for}%
\typeout{** the default language instead.}%
\else
\language=\csname l@#1\endcsname
\fi
#2}}
\providecommand{\BIBdecl}{\relax}
\BIBdecl

\bibitem{Deursen1998}
\BIBentryALTinterwordspacing
A.~V. Deursen and P.~Klint, ``Little languages: little maintenance?'' \emph{Journal of Software Maintenance: Research and Practice}, vol.~10, no.~2, p. 75–92, Mar. 1998. [Online]. Available: \url{http://dx.doi.org/10.1002/(SICI)1096-908X(199803/04)10:2<75::AID-SMR168>3.0.CO;2-5}
\BIBentrySTDinterwordspacing

\bibitem{schroder_grammars_2022}
\BIBentryALTinterwordspacing
M.~Schröder and J.~Cito, ``\BIBforeignlanguage{en}{Grammars for free: toward grammar inference for {Ad} {Hoc} parsers},'' in \emph{\BIBforeignlanguage{en}{Proceedings of the {ACM}/{IEEE} 44th {International} {Conference} on {Software} {Engineering}: {New} {Ideas} and {Emerging} {Results}}}.\hskip 1em plus 0.5em minus 0.4em\relax Pittsburgh Pennsylvania: ACM, May 2022, pp. 41--45. [Online]. Available: \url{https://dl.acm.org/doi/10.1145/3510455.3512787}
\BIBentrySTDinterwordspacing

\bibitem{knuth_semantics_1968}
\BIBentryALTinterwordspacing
D.~E. Knuth, ``\BIBforeignlanguage{en}{Semantics of context-free languages},'' \emph{\BIBforeignlanguage{en}{Mathematical Systems Theory}}, vol.~2, no.~2, pp. 127--145, Jun. 1968. [Online]. Available: \url{http://link.springer.com/10.1007/BF01692511}
\BIBentrySTDinterwordspacing

\bibitem{stevenson_survey_2014}
\BIBentryALTinterwordspacing
A.~Stevenson and J.~R. Cordy, ``\BIBforeignlanguage{en}{A survey of grammatical inference in software engineering},'' \emph{\BIBforeignlanguage{en}{Science of Computer Programming}}, vol.~96, pp. 444--459, Dec. 2014. [Online]. Available: \url{https://linkinghub.elsevier.com/retrieve/pii/S0167642314002469}
\BIBentrySTDinterwordspacing

\bibitem{gopinath_mining_2020}
\BIBentryALTinterwordspacing
R.~Gopinath, B.~Mathis, and A.~Zeller, ``\BIBforeignlanguage{en}{Mining input grammars from dynamic control flow},'' in \emph{\BIBforeignlanguage{en}{Proceedings of the 28th {ACM} {Joint} {Meeting} on {European} {Software} {Engineering} {Conference} and {Symposium} on the {Foundations} of {Software} {Engineering}}}.\hskip 1em plus 0.5em minus 0.4em\relax Virtual Event USA: ACM, Nov. 2020, pp. 172--183. [Online]. Available: \url{https://dl.acm.org/doi/10.1145/3368089.3409679}
\BIBentrySTDinterwordspacing

\bibitem{hoschele_mining_2016}
\BIBentryALTinterwordspacing
M.~Höschele and A.~Zeller, ``\BIBforeignlanguage{en}{Mining input grammars from dynamic taints},'' in \emph{\BIBforeignlanguage{en}{Proceedings of the 31st {IEEE}/{ACM} {International} {Conference} on {Automated} {Software} {Engineering}}}.\hskip 1em plus 0.5em minus 0.4em\relax Singapore Singapore: ACM, Aug. 2016, pp. 720--725. [Online]. Available: \url{https://dl.acm.org/doi/10.1145/2970276.2970321}
\BIBentrySTDinterwordspacing

\bibitem{moser_towards_2022}
\BIBentryALTinterwordspacing
M.~Moser, J.~Pichler, and A.~Pointner, ``Towards {Attribute} {Grammar} {Mining} by {Symbolic} {Execution},'' in \emph{2022 {IEEE} {International} {Conference} on {Software} {Analysis}, {Evolution} and {Reengineering} ({SANER})}.\hskip 1em plus 0.5em minus 0.4em\relax Honolulu, HI, USA: IEEE, Mar. 2022, pp. 822--826. [Online]. Available: \url{https://ieeexplore.ieee.org/document/9825876/}
\BIBentrySTDinterwordspacing

\bibitem{pointner_mining_2023}
A.~Pointner, ``Mining {Attributed} {Input} {Grammars} and their {Applications} in {Fuzzing},'' in \emph{2023 {IEEE} {Conference} on {Software} {Testing}, {Verification} and {Validation} ({ICST})}.\hskip 1em plus 0.5em minus 0.4em\relax Dublin, Ireland: IEEE, Apr. 2023, pp. 493--495.

\bibitem{pointner_generating_2025}
A.~Pointner, J.~Pichler, and H.~Prähofer, ``Generating {Inputs} for {Grammar} {Mining} using {Dynamic} {Symbolic} {Execution},'' \emph{The Art, Science, and Engineering of Programming}, vol.~10, no.~2, Jun. 2025, publisher: Aspect-Oriented Software Association (AOSA).

\bibitem{kalita_synthesis_2022}
\BIBentryALTinterwordspacing
P.~K. Kalita, M.~J. Kumar, and S.~Roy, ``\BIBforeignlanguage{en}{Synthesis of {Semantic} {Actions} in {Attribute} {Grammars}},'' Oct. 2022. [Online]. Available: \url{https://repositum.tuwien.at/handle/20.500.12708/81366}
\BIBentrySTDinterwordspacing

\bibitem{chen_towards_2017}
X.~Chen, C.~Liu, and D.~Song, ``Towards synthesizing complex programs from input-output examples,'' \emph{arXiv preprint arXiv:1706.01284}, 2017.

\bibitem{steinhofel_input_2022}
\BIBentryALTinterwordspacing
D.~Steinhöfel and A.~Zeller, ``\BIBforeignlanguage{en}{Input invariants},'' in \emph{\BIBforeignlanguage{en}{Proceedings of the 30th {ACM} {Joint} {European} {Software} {Engineering} {Conference} and {Symposium} on the {Foundations} of {Software} {Engineering}}}.\hskip 1em plus 0.5em minus 0.4em\relax Singapore Singapore: ACM, Nov. 2022, pp. 583--594. [Online]. Available: \url{https://dl.acm.org/doi/10.1145/3540250.3549139}
\BIBentrySTDinterwordspacing

\bibitem{kovacevic_grammar_2020}
\BIBentryALTinterwordspacing
Z.~Kova\u{c}evi\'{c}, M.~Mernik, M.~Ravber, and M.~\v{C}repin\u{s}ek, ``\BIBforeignlanguage{en}{From {Grammar} {Inference} to {Semantic} {Inference}—{An} {Evolutionary} {Approach}},'' \emph{\BIBforeignlanguage{en}{Mathematics}}, vol.~8, no.~5, p. 816, May 2020. [Online]. Available: \url{https://www.mdpi.com/2227-7390/8/5/816}
\BIBentrySTDinterwordspacing

\bibitem{mera_mining_2019}
\BIBentryALTinterwordspacing
M.~Mera, ``\BIBforeignlanguage{en}{Mining constraints for grammar fuzzing},'' in \emph{\BIBforeignlanguage{en}{Proceedings of the 28th {ACM} {SIGSOFT} {International} {Symposium} on {Software} {Testing} and {Analysis}}}.\hskip 1em plus 0.5em minus 0.4em\relax Beijing China: ACM, Jul. 2019, pp. 415--418. [Online]. Available: \url{https://dl.acm.org/doi/10.1145/3293882.3338983}
\BIBentrySTDinterwordspacing

\bibitem{zeller_fuzzing_2019}
\BIBentryALTinterwordspacing
A.~Zeller, R.~Gopinath, M.~Böhme, G.~Fraser, and C.~Holler, ``The fuzzing book,'' 2019. [Online]. Available: \url{https://www.fuzzingbook.org/}
\BIBentrySTDinterwordspacing

\bibitem{hodovan_grammarinator_2018}
\BIBentryALTinterwordspacing
R.~Hod\'{o}v\'{a}n, A.~Kiss, and T.~Gyim\'{o}thy, ``\BIBforeignlanguage{en}{Grammarinator: a grammar-based open source fuzzer},'' in \emph{\BIBforeignlanguage{en}{Proceedings of the 9th {ACM} {SIGSOFT} {International} {Workshop} on {Automating} {TEST} {Case} {Design}, {Selection}, and {Evaluation}}}.\hskip 1em plus 0.5em minus 0.4em\relax Lake Buena Vista FL USA: ACM, Nov. 2018, pp. 45--48, unread. [Online]. Available: \url{https://dl.acm.org/doi/10.1145/3278186.3278193}
\BIBentrySTDinterwordspacing

\bibitem{aho_compilers_2014}
A.~V. Aho, M.~S. Lam, R.~Sethi, and J.~D. Ullman, \emph{\BIBforeignlanguage{eng}{Compilers: principles, techniques, and tools}}, 2nd~ed., ser. Pearson custom library.\hskip 1em plus 0.5em minus 0.4em\relax Essex: Pearson, 2014.

\end{thebibliography}

\end{document}